\documentclass[aps,prl,showpacs,twocolumn,groupedaddress]{revtex4}

\usepackage{psfrag} 
\usepackage{graphicx}

\begin{document}
\title{Entanglement in the One-dimensional Kondo Necklace Model}
\author{A. \surname{Saguia}}
\email{asaguia@cbpf.br}
\affiliation{CBPF - Centro Brasileiro de Pesquisas F\'{\i}sicas \\
Rua Dr. Xavier Sigaud 150 - Urca, Rio de Janeiro, 22290-180,  Brazil.}
\author{M.S. \surname{Sarandy}}
\email{sarandy@dft.if.uerj.br}
\affiliation{UERJ - Universidade do Estado do
Rio de Janeiro\\
Rua S\~ao Francisco Xavier 524 - Maracan\~a,  \\ Rio de Janeiro, 20550-013,  Brazil.}

\date{\today}

\begin{abstract}

We discuss the thermal and magnetic entanglement in the one-dimensional Kondo
necklace
 model. Firstly, we show how the entanglement naturally present at
zero  temperature is distributed among pairs of spins according to the strength of the two
couplings of the chain, namely, the Kondo exchange interaction
and the  hopping energy. The effect of the temperature and  the presence of an
external magnetic field is then investigated, being discussed the adjustment
of these variables in order to control the entanglement
available in the system. In particular, it is indicated the existence of a
critical  magnetic field  above which the entanglement undergoes a sharp
variation, leading the ground state to a completely unentangled phase. 

\end{abstract}


\maketitle

\section{I. Introduction}

A striking feature in quantum mechanics is the presence of non-local 
correlations among different parts of a system. The physical property behind 
these purely quantum mechanical correlations is the so-called 
entanglement, which has been shown to be a key ingredient in quantum 
computation and communication \cite{r1}. Indeed,   
the search of experimental proposals to implement a quantum computer has 
strongly motivated the study of entanglement in condensed matter systems. In 
particular, schemes for quantum information processing based on Ising 
\cite{kane}, XY \cite{loss1}, and Heisenberg \cite{loss2}  
interactions have 
been suggested, leading to an intensive discussion about the thermal and 
magnetic entanglement in these models~\cite{ved1,wang,ved2,nielsen,amico}. 
Furthermore, the properties of entanglement have also been analysed in other 
condensed matter~\cite{delgado,ibose} and interacting quantum systems~\cite{plenio}.  

Recently, initial experiments towards the construction of quantum gates in  
rare-earth-ion doped crystals have been carried out \cite{kroll}. The
rare-earth compounds present several attractive properties for quantum
computation due to the existence of a partially filled $f$-shell, which is
shielded from  the environment by the outer electrons. Such partially filled
inner shell is a fundamental feature appearing in the heavy fermion
systems~\cite{r2}, characterized as metallic hosts in which magnetic
impurities are immersed. Thus, the discussion of the entanglement in
theoretical models describing the  heavy fermion physics turns out to be a
subject which deserves some attention. A standard hamiltonian for
this purpose is the Kondo lattice model \cite{kl,don}, 
which assumes the presence of one localized impurity spin
on each site, coupled to the conduction electrons. From the viewpoint of
the magnetic degrees of freedom, an important feature of the model is the 
competition between the intra-site Kondo screening of the localized magnetic
moment and the inter-site Ruderman-Kittel-Kasuya-Yosida (RKKY) interaction 
among the impurity spins, mediated by the conduction electrons. The Kondo
coupling favors a non-magnetic ground state, while the RKKY interaction tends
to establish a long range magnetic order. A scheme for entangling 
two conduction electrons in the presence of a magnetic impurity has been proposed  
by considering the Kondo hamiltonian in ref.~\cite{costa}.

A simplified version of the Kondo lattice hamiltonian was introduced by Doniach
in ref.~\cite{don}. This model, called Kondo necklace (KN), replaces, into the
Kondo interaction,
 the spins of the conduction electrons by a set of
pseudo-spins on a linear lattice,
 the charge degrees of freedom being frozen
out. In spite of this approximation, the interplay between the RKKY
interaction
 and the Kondo mechanism remains as an essential feature.
\begin{figure}  
 \includegraphics[angle=0,scale=0.45]{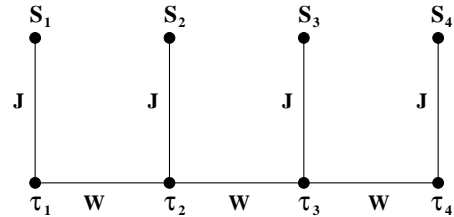}  
\caption{ \label{f1} Schematic arrangement of the KN chain
with  four sites.  The spin-1/2 operators $\tau^{\alpha}$ and $S^{\alpha}$,
where $\alpha=x,y,z$, denote the conduction eletrons and the spins of the
local
 moments, respectively.}
\end{figure}
The one-dimensional KN model is defined by the hamiltonian
\begin{eqnarray} 
H=W\sum^{N}_{i=1}({\tau}^{x}_{i}{\tau}^{x}_{i+1} +
{\tau}^{y}_{i}{\tau}^{y}_{i+1})
+  J \sum^{N}_{i=1} {\vec{S}}_{i} . {\vec{\tau}}_{i},
\label{knh} 
\end{eqnarray} 
where ${\tau}_{i}^\alpha$ and ${S}_{i}^\alpha$ are independent spin-1/2 operators
acting on a site $i$, which are given by $\sigma^{\alpha}/{2}$ , with
$\sigma^\alpha$ ($\alpha=x,y,z$) denoting the Pauli matrices.
The spin operators ${\tau}_{i}^\alpha$ and ${S}_{i}^\alpha$ are associated
with
 the conduction electrons and the innercore spins
hanging from the  $\tau$-spin chain, respectively. The hamiltonian
 Eq.~$(\ref{knh})$ 
describes a linear KN lattice with $N$ sites, where periodic
boundary conditions are adopted, {\it i.e.}, ${\tau}^{x}_{N+1}={\tau}^{x}_{1}$ and
${\tau}^{y}_{N+1}={\tau}^{y}_{1}$.
The positive parameter W represents the hopping energy and
J is the  Kondo exchange coupling, which can be either ferromagnetic ($J<0$)
or  antiferromagnetic ($J>0$). A schematic arrangement of the chain is
displayed in Fig.~\ref{f1}. Note that, since the KN model is
defined with two spins on each  site, the smallest representative cell of the
system is described by four  spins, each spin being taken as a quantum bit
(qubit).
A fully anisotropic version of the model, which we denote by X-KN,
will also be considered. In this case, the band of the conduction electrons is
represented just by an Ising  term $ W \sum
({\tau}^{x}_{i}{\tau}^{x}_{i+1})$. The isotropic KN
model, given in Eq.~$(\ref{knh})$, will be referred from now on as XY-KN.
   
The aim of this work is to discuss the thermal and magnetic entanglement 
present in the one-dimensional KN model.
 The paper is organized as follows.
In section II, we describe the natural thermal entanglement for both XY-KN and
X-KN models with two sites (four qubits).
 In section III, the entanglement in
a larger chain is analysed by considering the case of four sites (eight 
qubits). It is observed that, qualitatively, the entanglement of the system
keeps the main characteristics of the case with only two sites. Section IV
is devoted to the study of the effect of a magnetic 
field  tranversely applied to the $\tau$-chain. Finally, in section V we
summaryze our main results and present the
 conclusion.

\section{II. Thermal entanglement in a two-site KN ring}

In this section we study the entanglement between any two
qubits of the KN model with two sites at both zero and finite
temperature.
 Before describing our results we 
 present the definition of
concurrence, which is the
 measure of entanglement used throughout this paper.
The concurrence $C_{12}$ for a pair of qubits labelled as 1 and 2 is defined
as~\cite{concur}
\begin{equation}
C_{12} = \text{max}\left(\lambda_{1}-
\lambda_{2}-\lambda_{3}-\lambda_{4},0\right),
 \label{eq1}
\end{equation}
where the $\lambda_{i}$ are the square roots, in decreasing order, of the
eigenvalues of the operator 
\begin{equation}
R\equiv \rho_{12}(\sigma_{y}\otimes
\sigma_{y})\rho^{\ast}_{12}(\sigma_{y}\otimes\sigma_{y}).
\label{eq2}
\end{equation}
In Eq.~(\ref{eq2}), $\rho_{12}$ denotes the density matrix, which can be
either pure or mixed,  for the pair of qubits 1 and 2,
  and
$\rho^{\ast}_{12}$ its complex conjugate in the the standard basis
$\{|++\rangle,|+-\rangle,|-+\rangle,|--\rangle\}$. In a system with more than two
qubits, $\rho_{12}$ is obtained by tracing the density operator over the other qubits.
The concurrence ranges from 0, implying an unentangled state,
to 1, in which the two qubits are maximally entangled.

In pure states of $N$ qubits, it has been conjectured~\cite{cdist} that the entanglement
is distributed following the inequality for the squared concurrence
\begin{equation}
C_{12}^{2} +  C_{13}^{2} + ... + C_{1N}^{2} \le C_{1(23..N)}^{2},
\label{eq3}
\end{equation}
where $C_{1(23..N)}$ is the single-qubit concurrence, defined as
the concurrence between the qubit 1 and the rest of lattice $(23...N)$.
This quantity represents the collective contributions of the entanglement
between the qubit 1 and all the other qubits of the system and can be obtained
from $C_{1(23..N)} = 2\sqrt{\text{det}\,\rho_{1}}$, where $\rho_{1}$
is the density matrix for the qubit $1$. A relevant aspect of the single-qubit
concurrence is that, as it has been suggested in ref.~\cite{nielsen}, it may 
be a useful tool to identify a quantum critical point in a lattice system. 

\subsection{A. XY-KN MODEL}

Let us consider the entanglement for the pairs of qubits of the
two-site XY-KN model, whose spins are labelled with $\tau_{1}$, $\tau_{2}$,
$S_{1}$, and $S_{2}$, as
 represented in Fig.~\ref{f1}. For simplicity, we
will refer to $\tau_{1}$, 
 $\tau_{2}$, $S_{1}$, and $S_{2}$, as qubits $A$,
$B$, $C$, and $D$, respectively.
 The qubits $A$ and $B$ form a ring coupled
via antiferromagnetic
 XY interaction ($W>0$), while  $A$ and $C$, as well as
$B$ and $D$, are
 coupled via Heisenberg interaction, which can be either 
ferromagnetic  or antiferromagnetic. The symmetry of the lattice ensures that
the entanglement is the same for the following pairs: $AC$ and $BD$, $AD$ and
$BC$, and $AB$ and $CD$. Therefore we will only refer to the pairs $AB$, $AC$,
and $AD$ along the paper. 

The hamiltonian for the two-site model is obtained from Eq.~$(\ref{knh})$
by setting
 $N=2$.  The ground state of the system with either
$J<0$ or $J>0$ is nondegenerate and
can be written as the following pure state of $ABCD$: 
\begin{eqnarray}
|0\rangle_{xy} = N_{xy} \left( \frac{}{} |++--\rangle + |--++\rangle \right.
\nonumber \\
+\,\alpha_{1}(|+--+\rangle + |-++-\rangle) \nonumber \\
\left.+\,\alpha_{2}(|+-+-\rangle + |-+-+\rangle) 
\frac{}{}\right), \hspace{-0.48cm} \label{gsxy}
\end{eqnarray}
where the positions in the kets denote the qubits $A$, $B$, $C$, and $D$
in this order. The normalization constant $N_{xy}$ and the functions 
$\alpha_{1}$ and $\alpha_{2}$, which depend on the parameters $J$ and $W$, 
are found to be
\begin{eqnarray} 
\alpha_{1} = \frac{J + 2 \lambda_{xy}}{2 J}\,\,\, ,\,\,\, \alpha_{2} =
\frac{ \lambda_{xy}^2 + \lambda_{xy} J - 3 J^2 / 4}{WJ} \,\,,\hspace{0.9cm}
\nonumber \\  
N_{xy}=\frac{1}{\sqrt{2(1+\alpha_{1}^2+\alpha_{2}^2)}}\,,\hspace{4.1cm}  
\label{cxy} 
\end{eqnarray} 
with $\lambda_{xy}$ being the ground state energy 
\begin{eqnarray} 
\lambda_{xy} = -\frac{J}{6}-2\,\sqrt{Q}\,\,\text{cos}{\,\frac{\theta}{3}}\,,
\nonumber \hspace{3.9cm}\\ 
\theta = \text{arccos}
\left( \frac{-J(9W^2-16J^2)}{54\sqrt{Q^3}}\right)\, ,\, Q = \frac{4J^2 +
3W^2}{9}. \label{lxy}  
\end{eqnarray} 
Firstly we shall discuss the natural entanglement
at temperature T=0, for which only the ground state is populated. 
If the XY coupling is taken as zero ($W=0$), the system becomes two
decoupled Heisenberg dimers, namely, $AC$ and $BD$. In this regime,  all
the  pairs of qubits are completely unentangled for $J<0$. In fact, the 
absence of entanglement between qubits coupled by ferromagnetic Heisenberg
interaction  is a general property already pointed out in ref.~\cite{ved2}. On
the other hand, if $J>0$ then the qubits in the dimers are maximally entangled
for any value of the parameter $J$, with the concurrence vanishing for all the
other pairs. 

\begin{figure} 
\psfrag{C}{\hspace{-3.0cm}{{\Huge $C_{AB}$}}}
\psfrag{W}{\hspace{-2.0cm}{\Huge $W$}}
\psfrag{J}{\hspace{1.7cm}\Huge $J$}
\centering 
{\includegraphics[angle=0,scale=0.35]{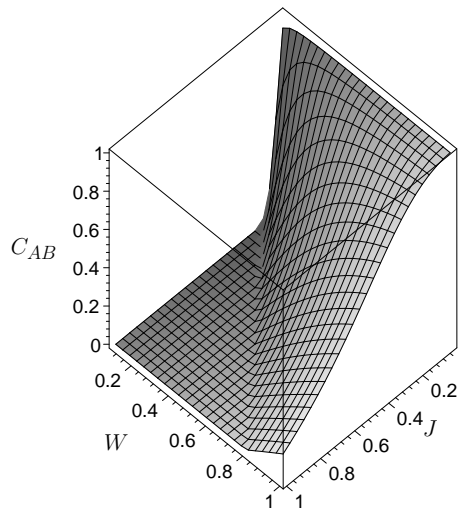}}
\caption{ \label{f2}  Concurrence 
between the qubits $A$ and $B$ in the two-site XY-KN model for
antiferromagnetic Heisenberg interaction at zero temperature.} 
\end{figure}

\begin{figure}  
\psfrag{C}{\hspace{-3.0cm}{{\Huge $C_{AC}$}}}
\psfrag{W}{\hspace{-2.0cm}{\Huge $W$}}
\psfrag{J}{\hspace{1.7cm}{\Huge $J$}}
\centering  
{\includegraphics[angle=0,scale=0.35]{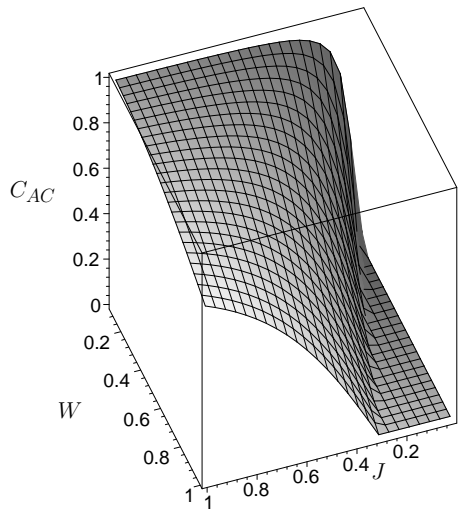}}
\caption{ \label{f2l}  Concurrence 
between the qubits $A$ and $C$ in the two-site XY-KN model for
antiferromagnetic Heisenberg interaction at zero temperature.}
\end{figure}

Consider now the parameter $W>0$. We obtain, in the limit $J=0$,
the two-qubit antiferromagnetic XY ring in the presence of two other
non-interacting spins ($C$ and $D$). In this case, the pair of qubits $AB$ is
maximally entangled for any $W$ and the concurrences for the other pairs are
zero. For the Heisenberg coupling $J>0$,
there is a competition,  governed by the values of
$J$ and $W$,  between the entanglement of the pairs $AB$ and $AC$ coupled by 
Heisenberg and XY interactions, respectively. The 
concurrences $C_{AB}$ and $C_{AC}$  
are shown in Figs.~\ref{f2} and~\ref{f2l}, the pair $AD$
 being always
unentangled. By fixing the value
 of $W$, it is apparent  that a large value
of $J$ favors the
 entanglement between $AC$, while small $J$ tends to increase
the entanglement between the qubits $A$ and $B$. 
In fact, the concurrences $AB$ and $AC$ are complementary, {\it i.e.}, if  the
qubit $A$ is partially entangled with $B$, then $A$ can have only a limited
quantity of entanglement with $C$.  
These results satisfy completely the inequality (\ref{eq3}), with  
the upper bound limit for the entanglement being obtained
from the single-qubit concurrence between $A$ and $BCD$, given by 
$C_{A(BCD)}=1$ independently of the  values of $J$ and $W$.

\begin{figure}  
\psfrag{C}{\hspace{-3.0cm}{{\Huge $C_{AB}$}}}
\psfrag{W}{\hspace{-2.0cm}{\Huge $W$}}
\psfrag{J}{\hspace{0.8cm}{\Huge $J$}}
\centering  
{\includegraphics[angle=0,scale=0.35]{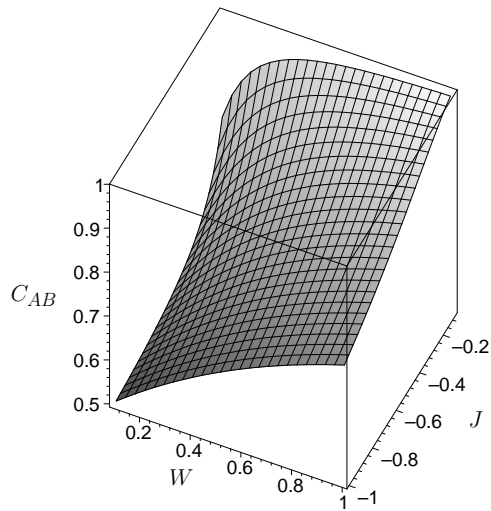}}
\caption{ \label{ff}  Concurrence  
between the qubits $A$ and $B$ in the two-site XY-KN model for ferromagnetic
Heisenberg interaction at zero temperature.}  
\end{figure}

\begin{figure} 
\psfrag{C}{\hspace{-2.0cm}{{\Huge $C_{AD}$}}}
\psfrag{W}{\hspace{-2.0cm}{\Huge $W$}}
\psfrag{J}{\hspace{1.0cm}{\Huge $J$}}
\centering
{\includegraphics[angle=0,scale=0.35]{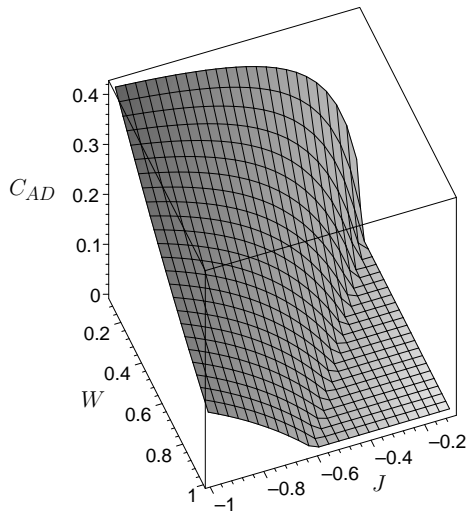}} 
\caption{ \label{ffl}  Concurrence  
between the qubits $A$ and $D$ in the two-site XY-KN model for ferromagnetic
Heisenberg interaction at zero temperature.}  
\end{figure}

For the case $J<0$, the entanglement is distributed between the pairs $AB$ 
and $AD$, as plotted in Figs.~\ref{ff} and~\ref{ffl}. We can see that, 
although there is no
concurrence between qubits coupled with Heisenberg interaction,
an increase of the value of $J$ decreases the entanglement of the pair $AB$
and increases the concurrence for $AD$. In fact, it can be observed
from Eqs.~(\ref{gsxy}-\ref{lxy}), that  a variation in $J$ affects
$C_{AB}$ and $C_{AD}$, since the functions $\alpha_{1}$ and $\alpha_{2}$ depend
on this parameter. As in the case $J>0$, the inequality (\ref{eq3}) is
satisfied with $C_{A(BCD)}=1$.

\begin{figure} 
\psfrag{C}{\hspace{-3.0cm}{{\Huge $C_{AB}$}}}
\psfrag{J}{\hspace{1.5cm}{\Huge $J$}}
\psfrag{kT}{\hspace{-2.5cm}{\Huge $kT$}}
\centering 
{\includegraphics[angle=0,scale=0.35]{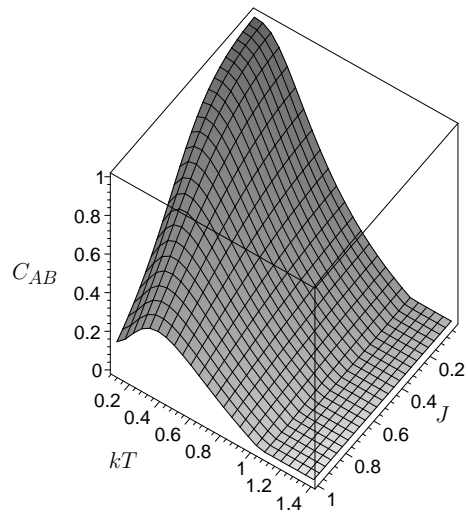}} 
\caption{ \label{f3}  Thermal concurrence in the two-site XY-KN model
with antiferromagnetic Heisenberg interaction for the pair $AB$. 
We have set the parameter $W$ to $1$.}    
\end{figure}

\begin{figure}
\psfrag{C}{\hspace{-2.0cm}{{\Huge $C_{AC}$}}}
\psfrag{J}{\hspace{1.7cm}{\Huge $J$}}
\psfrag{kT}{\hspace{-2.5cm}{\Huge $kT$}}
\psfrag{0.1}{\,}
\psfrag{0.3}{\,}
\psfrag{0.5}{\,}
\centering
{\includegraphics[angle=0,scale=0.35]{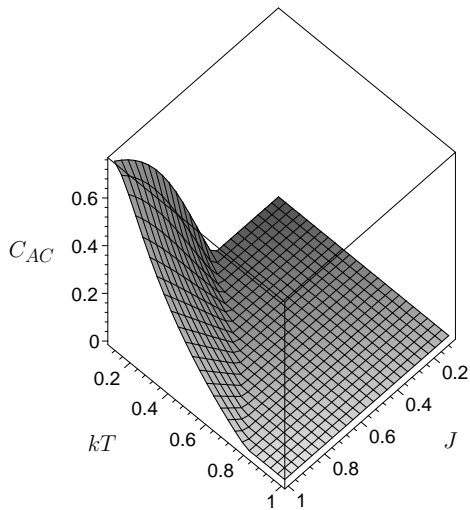}} 
\caption{ \label{f3l}  Thermal concurrence in the two-site XY-KN model
with antiferromagnetic Heisenberg interaction for the pair $AC$. 
We have set the parameter $W$ to $1$.}  
 
\end{figure}

In order to take into account the effect of the temperature on
the entanglement in the XY-KN model, we consider the density matrix for
the system at thermal
 equilibrium as given by $\rho(T)=exp(-H/kT)/Z$, where Z
is the partition
 function and $k$ the Boltzmann's constant. 
In
Figs.~\ref{f3} and~\ref{f3l} we show the concurrences 
for the pairs of qubits $AB$ and $AC$ as functions of the temperature for an
antiferromagnetic Heisenberg interaction and with fixed $W=1$. By comparing the
Fig.~\ref{f2} to~\ref{f3} and
 Fig.~\ref{f2l} to~\ref{f3l} we can see that,
for the regime of very low temperatures, the distribution of entanglement for
the pairs of qubits at $T=0$ is recovered. As the temperature increases,
 the
entanglement for the pair $AB$ increases, for all values of $J$, before
beginning to decrease. This is explained by observing that, for $J=0$, the
ground state is degenerate, being a statistical mixture of four states
with energy eigenvalue $-W$. In this case, the 
entanglement $AB$ assumes the maximum value $1$ at $T=0$, independently of the
parameter $W$. When $J$ is "turned on", the degeneracy splits up and the ground
state turns out to be given by Eq.~(\ref{gsxy}). However, for a nonvanishing
value of the coupling $J$, there are some excited states contributing with a 
greater portion of entaglement $AB$ than the ground state. Thus, since
the effect of the temperature is to populate the excited states, the
entanglement $AB$ turns out to increase before being completely destroyed.
In contrast, the entanglement for the pair $AC$ always decreases. We can
understand this behavior by noting that, for $W=0$, the ground state of the
system is nondegenerate and the pair $AC$ is maximally entangled for any value
of $J$ at $T=0$. For $W$ nonvanishing, the ground state of the system
remains nondegenerate and there are no excited states contributing with a
quantity of entanglement greater than the ground state. Therefore, the
mixing of Eq.~(\ref{gsxy}) with higher energy levels, in this case, can only
destroy the entanglement. Moreover, as greater is the coupling $J$, greater
has to be the temperature to lead $C_{AC}$ to 0. 

For ferromagnetic Heisenberg
interaction, the thermal concurrence is different from zero for the
pairs $AB$ and $AD$. For $AB$ it is possible to increase the
entanglement with the temperature. However, the quantity of entanglement
generated is very small compared to the case of the antiferromagnetic
Heisenberg interaction displayed in Fig.~\ref{f3}.   

\subsection{B. X-KN MODEL}

\begin{figure}   
\psfrag{C}{\hspace{-3.0cm}{{\Huge $C_{AC}$}}}
\psfrag{W}{\hspace{-2.0cm}{\Huge $W$}}
\psfrag{J}{\hspace{1.7cm}{\Huge $J$}}
\centering
\includegraphics[angle=0,scale=0.35]{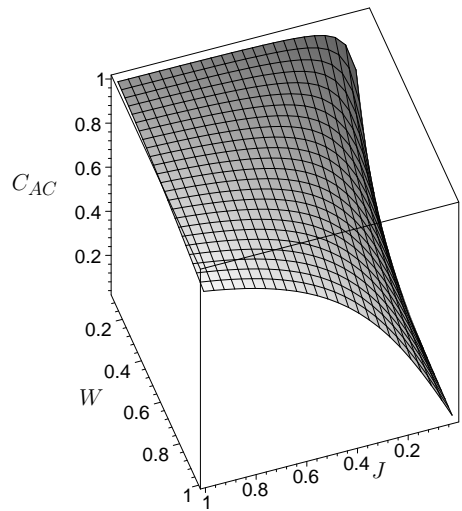}    \caption{\label{f4} The
concurrence between the qubits $A$ and $C$ in the two-site X-KN model.
The concurrences for the pairs $AB$ and $AD$ are  zero.}   
\end{figure}

In the anisotropic version of the KN model, the $\tau$-chain
 displayed 
in Fig.~\ref{f1} is composed by a set of qubits coupled via Ising
interactions.
 As it is well known~\cite{ved1,nielsen,amico}, there is no
entanglement naturally present for any pairs of qubits in the pure Ising
chain. However, when a spin-$S_i$ is coupled to a spin-$\tau_i$ of
the Ising chain, by means of an antiferromagnetic Heisenberg interaction, the
pair of qubits $\tau_{i}S_{i}$ becomes immediately entangled. It is worth 
reminding that ferromagnetic Heisenberg coupling cannot induce any concurrence
in the system. In order to  describe in detail the entanglement in the two-site
X-KN model, let us consider four qubits $A$, $B$, $C$, and
$D$, firstly at zero temperature. Similarly as defined for XY-KN model, the
pair $AB$ is coupled by an Ising interaction $W>0$ and the pairs $AC$ and $BD$
by an antiferromagnetic Heisenberg interaction $J$. The ground state of the
system can be written as 
\begin{eqnarray} 
|0\rangle_{x} = N_{x} \left( \frac{}{} |++++\rangle +
|----\rangle \right. \nonumber 
\\ -\, \beta (|++--\rangle + |--++\rangle) \nonumber
\\ +\, \beta (|+--+\rangle + |-++-\rangle) \nonumber \\
\left.-\,(|+-+-\rangle + |-+-+\rangle) 
\frac{}{}\right), \hspace{-0.3cm} \label{gsx}
\end{eqnarray} 
where the function $\beta$ and the normalization constant $N_{x}$, which
depend  on the parameters $J$ and $W$, are found to be 
\begin{eqnarray}
\beta = \frac{2J+\sqrt{4J^2+W^2}}{W}
\,\,\, ,\,\,\,
N_{x}=\frac{1}{\sqrt{4(1+\beta^2)}}\,. \label{cx}
\end{eqnarray}
From Eqs.~(\ref{gsx}) and (\ref{cx}), it can be shown that the entanglement for
both pairs $AB$ and $AD$ is zero. The concurrence for the pair $AC$
can be expressed as 
\begin{equation}
C_{AC}=\frac{\beta^2 - 1}{\beta^2 + 1}\, , \label{cacx}
\end{equation}
reaching the maximum value in the limit 
$W=0$.  In Fig.~\ref{f4}, the concurrence $C_{AC}$ is plotted as a function of
the parameters  $J$ and $W$.
 As it can be observed, although the coupling
$W$ cannot induce any entanglement for the Ising coupled pairs, it contributes
to decrease $C_{AC}$.

The single-qubit concurrence
 $C_{A(BCD)}$, between $A$ and the
rest of the chain ($BCD$), has also been computed. 
For $J=0$, we obtain $C_{A(BCD)}=0$ and, for any nonvanishing $J>0$, we find
$C_{A(BCD)}=1$ independently of the parameter $W$. This
result indicates that, in $J=0$, the system undergoes a quantum phase
transition~\cite{sachdev}
 characterized by a jump in the entanglement of the
qubit $A$ with $BCD$, from 0 to 1, 
 for an infinitesimal increase of $J$.
 In
other words, it happens a  fundamental modification in the pattern of
entanglement present in the ground state, which changes from a completely
unentangled state to one in which the upper bound limit for the concurrence
between any two qubits assumes the maximal value. This quantum phase
transition is very similar to that found in ref.~\cite{ved1} for the
transverse magnetic field Ising chain, with the   
coupling $J$ playing the role of the transverse field.
Indeed, in the presence of random coupling constants, it has been pointed out
in ref.~\cite{a1} a parallelism of the critical magnetic properties between the
X-KN and transverse Ising models. The results found here for the
X-KN system indicate, in the pure case (fixed $J$ and $W$), a 
similarity in the behavior of the entanglement of both models. 

\begin{figure}  
\psfrag{C}{\hspace{-2.0cm}{{\Huge $C_{AC}$}}}
\psfrag{J}{\hspace{1.7cm}{\Huge $J$}}
\psfrag{kT}{\hspace{-2.0cm}{\Huge $kT$}}
\centering
\includegraphics[angle=0,scale=0.35]{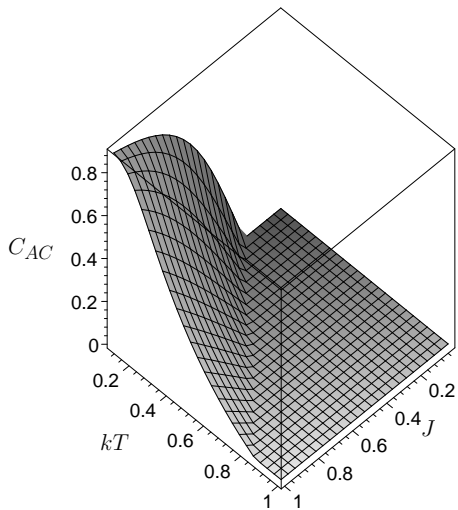}  
\caption{ \label{f5}Thermal concurrence for the
pair of qubits $AC$ in the two-site X-KN model. We have set the parameter $W$
to $1$.}  
  \end{figure}

For completeness, we show in 
Fig.~\ref{f5}  the entanglement of the pair $AC$ as a funtion of temperature, 
where we have set the 
 parameter $W=1$.
We observe that the temperature tends always to decrease the entanglement.

\section{III. The entanglement for a larger chain}

In this section, we numerically 
describe the entaglement in the KN model with a larger number of sites. 
By  considering the chain  with four sites, as represented in
Fig.~\ref{f1}, we show that the results found for the
entanglement are qualitatively very
 close to that
obtained earlier for a two-site X-KN
 and XY-KN ring. In fact, this 
seems to rest on the competition between the two couplings
of the theory,  which will take place independently of the number of sites
considered.

\begin{figure} 
\centering 
{\includegraphics[angle=0,scale=0.35]{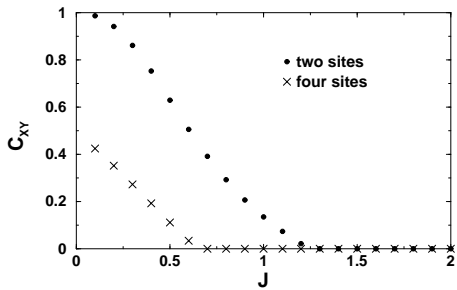}} 
\caption{ \label{f61}  Concurrence at $T=0$ for 
neighbor $\tau$-spins in the XY-KN model with two
and four sites. We have set $W=1$ and $J>0$.} 
 \end{figure}

\begin{figure}  
\centering  
{\includegraphics[angle=0,scale=0.35]{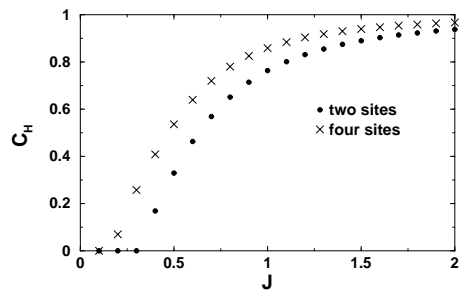}}  
\caption{ \label{f62} Concurrence at $T=0$ for  
the pairs of qubits coupled by antiferromagnetic Heisenberg interaction in the
XY-KN model with two and four sites. We have set $W=1$.}    
\end{figure}

Here we discuss the four-site XY-KN model with
antiferromagnetic Heisenberg coupling at $T=0$. In this case, the
concurrence is distributed among the Heisenberg coupled qubits and the neighbor
and next-neighbor pairs in the $\tau$-chain. In Figs.~\ref{f61}
and~\ref{f62} we compare the entanglement by taking the
chain with two and four sites. We call $C_{XY}$  the concurrence
 between
neighbor spins on the $\tau$-chain,  and  $C_{H}$, the concurrence
 for
neighbor spins coupled by Heisenberg interaction.
 We observe that, for a
fixed coupling $W$, the concurrence $C_{XY}$ decreases faster with the rise
of $J$ as the number of sites becomes larger. On the other hand, $C_H$ is
favored if many sites are taking into account, increasing faster with $J$ as we
increase the size of the chain. The single-qubit concurrence remains 1 and the
inequality (\ref{eq3}) is still obeyed.  


In the case of a chain with an odd number of sites we found that, due to the 
breaking of the ground state translational invariance~\cite{w2}, the entanglement
between neighbor qubits on the $\tau$-chain can be different from
each other. However, this
 difference in the behavior of odd and even number
of sites is expected to disapear as the size of the chain becomes very large,
as discussed for the Heisenberg model in ref.~\cite{ved2}.

\section{IV. Magnetic field along the z axis}

We investigate here how an external magnetic field
applied in the KN model along the z direction can modify the entanglement of
the
 system. The hamiltonian for this case
 is given by
\begin{eqnarray} 
H_{B}= H + B\sum^{N}_{i=1}({s}^{z}_{i} +{\tau}^{z}_{i}), 
\label{knhb} 
\end{eqnarray} 
where $H$ is the hamiltonian previously defined in Eq.~(\ref{knh}) and $B$
is a
 constant magnetic field. 

In order to describe the magnetic entanglement in the two-site
XY-KN and X-KN  models, we define qubits A, B, C and D in the same
way as  before. As discussed in section III, the qualitative results found for
the two-site system hold for a larger chain, leading us to be concerned here in
the former case. We first consider the magnetic entanglement in the KN model
with antiferromagnetic Heisenberg coupling at zero temperature. For
$B=0$, the ground states of the XY-KN and X-KN models are 
given by Eqs.~(\ref{gsxy}) and (\ref{gsx}), respectively. 
The single-qubit
concurrence  in both cases is always 1, independently of the values of the
parameters $J$ and $W$. After the external magnetic field $B$ is
 "turned on", 
the ground state of the XY-KN model suddenly changes, as $B$ crosses a
specific value, from Eq.~(\ref{gsxy}) to 
\begin{eqnarray}
|0\rangle^{B}_{xy}=N^{B}_{xy}\left( \frac{}{} \gamma_{1}(|---+\rangle -
|--+-\rangle)  \hspace{1.5cm} \right. \nonumber \\  \left. -
\gamma_{2}(|-+--\rangle + |+---\rangle)\frac{}{} \right) \hspace{1cm} 
\label{gsxysb}
\end{eqnarray}
where $N^{B}_{xy}$ is the normalization constant and the $\gamma_{i}$
are functions of the parameters $J$ and $W$. In this state the
single-qubit concurrence has a little decrease from 1 to a value which
depends on $J$ and $W$. For example, when $J=W=1$, the value of the
single-qubit concurrence is $0.989$. By increasing the magnetic field to a 
critical value $B_{c}=B_{c}(J,W)$ the single-qubit
concurrence vanishes, since the ground state becomes the fully
unentangled state $|----\rangle$. Hence, for any fixed value of the parameter
$W$,  we obtain a critical line $B_c = B_c (J)$. Taking $J=W$, we can 
numerically interporlate the straight line $B_c \approx 1.707 \, J$. We observe that
the effect of the external magnetic field is the gradual alignement of all the
spins in the chain, leading the ground state always to an unentangled state
for a high value of  $B$.    

\begin{figure} 
\centering 
{\includegraphics[angle=0,scale=0.35]{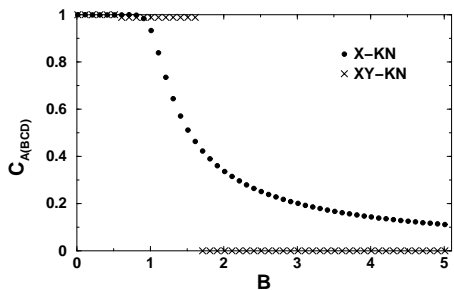}} 
\caption{The single-qubit concurrence for both two-site XY-KN and X-KN models
with antiferromagnetic Heisenberg interaction at T=0. The parameters $J$ and
$W$ are set to 1.
 For the XY-KN model the system presents a quantum phase
transition at $B_{c} \approx 1.707$.
 For the X-KN model the concurrence 
decays analytically with $B$.}  \label{f7} \end{figure}

The behavior for the X-KN model in the presence of a
magnetic field is quite different to that of the XY-KN model. When we "turn 
on" the magnetic field, the ground state given by
Eq.~(\ref{gsx}) generalizes to 
\begin{eqnarray}
|0\rangle^{B}_{x} = N^{B}_{x} \left( \frac{}{} \epsilon_{1}|++++\rangle +
\epsilon_{2}|----\rangle \right. \nonumber 
\\ +\, \epsilon_{3}|++--\rangle + \epsilon_{4}|--++\rangle \nonumber
\\ +\, \epsilon_{5} (|+--+\rangle + |-++-\rangle) \nonumber \\
\left.+\,\epsilon_{6}(|+-+-\rangle + |-+-+\rangle) 
\frac{}{}\right), \hspace{-0.45cm} \label{gsxsb}
\end{eqnarray}
where $N^{B}_{x}$ is the normalization constant and the $\epsilon_{i}$
are functions of the parameters $J$ and $W$.
By increasing $B$, all the functions $\epsilon_{i}$
decrease except $\epsilon_{2}$, which becomes higher. In the limit $B
\rightarrow \infty$ the ground state becomes the unentangled state 
$|----\rangle$.  In Fig.~\ref{f7} it is exhibited the single-qubit concurrence
for the XY-KN and X-KN models at zero temperature, with the parameters $J$ and
$W$ set to one. As it can be seen, 
the single-qubit concurrence is an analytical function of the magnetic field
$B$ for the  X-KN model, showing a smooth decrease of the entanglement of the
ground state. This behavior is clearly different from the XY-KN case, in
which the concurrence undergoes a sharp variation. 

\begin{figure}  
\centering  
{\includegraphics[angle=0,scale=0.35]{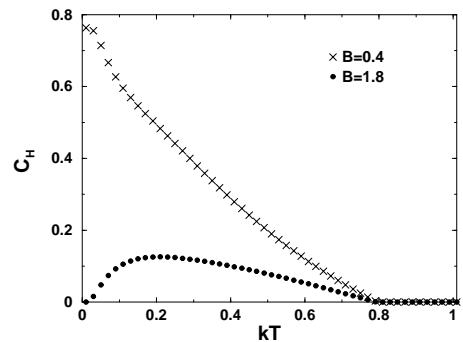}}  
\caption{Thermal concurrence for the pairs of
qubits coupled by antiferromagnetic Heisenberg interaction in the two-site
XY-KN model. The parameters $J$ and $W$ are set to
$1$. As $B$ crosses the critical value $B_c \approx 1.707$ the concurrence 
increases with the temperature before vanishing at $T \approx 0.79$.} 
\label{f8}  \end{figure}

Let us now consider the thermal entanglement in the model by taking into
account the  presence of the magnetic field $B$. As discussed in Section $IIA$
for the XY-KN chain with antiferromagentic Heisenberg interaction, the increase
of the temperature can rise the entanglement of $AB$ due to the mixing of the
ground state with some higher entangled excited states. In contrast, the
concurrence for the pair $AC$ is always decreased as the temperature is
"turned on". However, when  a magnetic field $B$ greater than the critical
value $B_c$ is applied in the system, the concurrence for the pair $AC$ can
considerably be increased as the temperature rises, as exhibited in
Fig.~\ref{f8}. This is because, as $B$ crosses $B_c$, the ground state of the
two-site XY-KN model suddenly changes from a highly entangled to an
unentangled state. Hence, the increase of the temperature can here generate
entanglement also for the pair $AC$ due to the mixing of the unentangled
ground state with some entangled excited states. For $B > B_c$, as $B$
approaches the critical field $B_c$,  the entanglement becomes higher with the
increase of the temperature.  Moreover numerical simulations point to the
existence of a temperature for which the entanglement is always destroyed
indenpendently of the magnetic field applied. For $J=W=1$, this temperature is
$T_{c} \approx 0.79$ when we compute the concurrence for a Heseinberg coupled
pair. 

We have also studied the behavior of the concurrence in the
case of ferromagnetic Heisenberg coupling in the presence of the transverse
magnetic field. The results found for the single-qubit concurrence are
qualitatively the same as the antiferromagnetic case in both X-KN and XY-KN
models, the difference being that now a smaller value of $B$ is necessary for
destroying the entanglement. 

\section{V. Conclusion} 

In this paper we have discussed the thermal and magnetic
entanglement in the one-dimensional KN hamiltonian, which is a
simple model for  describing heavy fermion systems.  
Concerning the natural entanglement present in the model, we have found that
the concurrence for a pair of qubits is distributed according to the
fundamental couplings of theory. Furthermore we have been able to indicate, 
in the fully anisotropic version of the system, {\it i.e.} the X-KN model, a 
quantum phase transition for the Heisenberg interaction $J=0$.  The
critical behavior of the entanglement has been characterized by means of the
single-qubit concurrence, which is the upper bound limit for the entanglement
present in the chain. The effect of the temperature and the presence of an
external magnetic field applied in the z direction has also been studied. 
These variables have been shown to be adjustable to control the entanglement
available in the system. In particular, it has  been found, in the XY-KN
model, a critical magnetic field $B_{c}(J,W)$ which separates
phases exhibiting a completely different pattern of entanglement. 

It is known the KN model presents a rich phase diagram with remarkable
critical magnetic properties \cite{a2}. It would be interesting to 
search a connection of this critical behavior with the results found in this
paper. Efforts in this direction would contribute to a more complete
understanding of the relationship between entanglement and quantum phase
transitions in condensed matter systems as originally proposed in
refs.~\cite{nielsen,amico}.

\section{Acknowledgements}

We would like to
 thank Conselho Nacional de De\-sen\-vol\-vi\-men\-to
Cient{\'{\i}}fico
 e 
Tecnol\'ogico-CNPq Brasil  for  financial 
support.

\end{document}